\documentclass[prl,final,twocolumn,showpacs,byrevtex,float]{revtex4}
\usepackage{graphicx}
\usepackage{epsfig}
\begin{document}
\def\BibTeX{\rm B{\sc ib}\TeX}

\title{Symmetry of standing waves generated by a point defect in epitaxial graphene\\}
\author{L. Simon$^{1}$ $\footnote[1]{corresponding author \\Email address:
L.Simon@uha.fr}$, C. Bena$^{2}$, F. Vonau$^{1}$, D. Aubel$^{1}$,
H. Nasrallah$^{1}$, M. Habar$^{1}$, J. C. Perruchetti$^{1}$}
\affiliation{$^{1}$Laboratoire de Physique et de Spectroscopie Electronique\\
CNRS-UMR7014, 4, rue des Fr\`eres Lumi\`ere 68093
Mulhouse-France\\$^{2}$ Institut de Physique Th\'eorique, CEA/Saclay, CNRS-URA 2306\\
Orme des Merisiers, F-91191 Gif-sur-Yvette, France}

\date{\today}

\begin{abstract}
Using scanning tunneling microscopy (STM) and Fourier Transform
STM (FT-STM), we have studied a point defect in an epitaxial
graphene sample grown on silicon carbide substrate. This analysis
allows us to extract the quasiparticle energy dispersion, and to
give a first experimental proof of the validity of Fermi liquid
theory in graphene for a wide range of energies from -800 $meV$ to
+800 $meV$. We also find evidence of a strong threefold anisotropy
in the standing waves generated by the defect. We discuss possible
relations between this anisotropy, the chirality of the electrons,
and the asymmetry between graphene's two sublattices. All
experimental measurements are compared and related to theoretical
T-matrix calculations.
\end{abstract}

\pacs{68.65.-k, 81.16.Fg, 81.07.-b, 81.16.Rf, 82.30.RS, 82.65.+r}

\maketitle

Graphene is a two-dimensional zero-gap semiconductor that shows
many fascinating physical properties, in particular the
quasiparticles near the Dirac point have a massless spectrum and
linear dispersion. These relativistic particles have a velocity of
about one percent of the speed of light, and are able to propagate
ballistically over mesoscale distances \cite{NovoselovNature2005}.
The two inequivalent A and B sublattices of the 2D honeycomb
carbon structure give rise to the fermionic pseudospin and
chirality that lie at the origin of the forbidden backscattering
in the presence of extended impurities \cite{falko}, and of the
anomalous Friedel oscillations in the vicinity of localized
impurities \cite{BenaPRL08}.
The possibility to make few-layer graphene samples by the
annealing of a silicon carbide substrate has been previously
demonstrated in Refs.~\cite{ForbeauxPRB98,SimonPRB99,others}. The
Dirac fermion behavior has been evidenced in epitaxial graphene by
angle-resolved photoemission spectroscopy (ARPES)
\cite{BostwickNaturephysics07,ZhouNaturePhysics07}.

Here we analyze the Fourier transform of STM conductance images in
bilayer epitaxial graphene. It has been shown that such a study
should allow one to extract the quasiparticle dispersion. This was
first observed in superconductors \cite{McElroyNature03}, then by
some of the authors in a semi-metallic material
\cite{VonauPRL05,SimonJcondMat07}, and more recently in bilayer
graphene \cite{RutterScience07}; the latter experiment was however
limited to a small energy window ($\pm 100 meV$). Various other
STM observations of the LDOS distortions in the presence of
impurities in few-layer graphene exist in the literature (see for
example
\cite{RuffieuxPRB05,RuffieuxPRL00,KellySurfSc98,NiimiPRL06}, for
hydrogen-induced defects on graphite surfaces, and
\cite{RutterScience07,MalletPRBR07} for several types of defects
in graphene bilayers).

In this Letter we present an in-depth theoretical (single impurity
T-matrix) and experimental (STM and FT-STM) study of the local
density of states (LDOS) in the vicinity of an impurity. We have
looked for point defects and we have focused on  a single isolated
impurity that strongly affects the LDOS of the epitaxial graphene
layer; the distortion induced by this defect turns out to have a
threefold symmetry. Using FT-STM we test the robustness of the
quasiparticle dispersion, and the range of validity of the Fermi
liquid theory in bilayer (or multilayer) graphene for a large
energy window ($-800$ to $+800$ $meV$). Our experiment was
performed with a LT-STM from Omicron at 77K in a base pressure of
$10^{-11}$ mbar. The samples were prepared in UHV by the annealing
of n-doped SiC(0001) at 1000K starting from a 3x3 reconstruction
\cite{SimonPRB99}.

\begin{figure}
\begin{center}
\includegraphics[width=8cm]{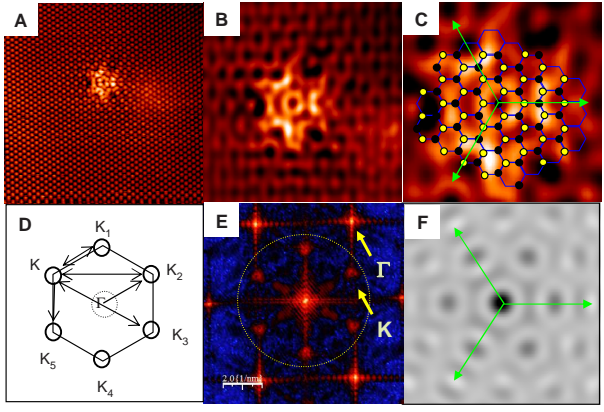}\\
\end{center}
\caption{A) STM topographic image ($10\times 10 nm^{2}$, -17$meV$,
1$nA$) showing an isolated defect. B) Zoom-in ($5\times 5 nm^{2}$)
on the defect, subsequent to a FFT filtering removing the atomic
resolution and all other features with wavevectors outside the
yellow circle in E. C) $2.5\times 2.5 nm^{2}$ zoom in on B with
the schematic atomic lattice of graphene (A/black and B/yellow
atoms) superimposed over the standing-wave pattern around the
point defect. D) schematic representation of the 2D Brillouin zone
and of the constant energy contours, indicating the possible
scattering processes. E) The FFT power spectrum of the 2D
topographic image in A. F) The real-space LDOS in a bilayer
graphene calculated for a point defect using a single-impurity
T-matrix approximation at $300meV$ above the Dirac
point.}\label{Fig1}
\end{figure}

Figure \ref{Fig1}A shows a topographic image of a large graphene
terrace taken at -17 $meV$ (probing full states). This layer shows
an intriguing ``star-like'' defect with an apparent six-fold
($C6v$) symmetry. This atomic defect is accompanied by a strong
distortion of the graphene lattice. The center appears black,
which is a dramatic change from the case of the unperturbed
lattice.  As schematized in figure \ref{Fig1}C, a detailed
analysis of the real-space image, shows that the point defect is
directly located over or under a lattice site. The 2D Fourier
transform of the topographic image is shown in \ref{Fig1}E. Six
high-intensity regions corresponding to the periodicity of the
hexagonal lattice appear at the centers $\Gamma$ of the six
Brillouin zones (BZ) neighboring the first BZ. Six high-intensity
features (open circles) with an anisotropic intensity are also
observed around the corners of the first BZ. They can be
attributed to the scattering of quasiparticles between isocontours
surrounding the K points (the corners of the BZ) as schematized in
figure \ref{Fig1}D. This type of scattering is called inter-nodal
or inter-valley scattering. The six high-intensity regions at the
corners of the first BZ correspond to scattering processes that
link two consecutive isocontours. Similar features have also been
observed in
\cite{RuffieuxPRL00,RutterScience07,MalletPRBR07,malletnew}.

Using FFT filtering \cite{FFTProcedure} we have removed the
lattice periodicity vectors and all other features with
wavevectors outside the yellow circle in Fig.~\ref{Fig1}E, thus
taking into account only the intra-valley and the
inter-consecutive-valley scattering processes ($K\rightarrow
K_{1}$ type) depicted in figure \ref{Fig1}D. The resulting
real-space image is shown in figure \ref{Fig1}B. This operation
strongly enhances the anisotropic intensity observed also on the
bare topographic image. The LDOS near the defect shows a clear
threefold ($C3v$) symmetry. While, as depicted in
Fig.~\ref{Fig1}C, close to the impurity we observe a fairly
homogeneous standing-wave ring, with an almost perfect six-fold
symmetry, farther away from the impurity, the intensity is clearly
higher along three axes (drawn in green in Figs.~\ref{Fig1}C and
\ref{Fig1}F). direction.

Let us now analyze the possible origin of this three-fold
symmetry. If the defect is localized above an A atom (shown in
black in Fig.~\ref{Fig1}C), the nearest neighbors are three B
atoms ($C3v$ symmetry). The second-nearest neighbors are six A
atoms ($C6v$ symmetry). Hence, the three-fold symmetry comes from
the lattice environment, in particular from the scattering of
electrons between the A and the B sublattices. This is related to
the pseudospin and chirality of electrons in graphene and will be
explained in more detail in what follows.

The theoretical LDOS modulation for a bilayer graphene in the
presence of a single impurity is represented in Fig.~\ref{Fig1}F.
This was obtained at an energy of $300meV$ above the Fermi level,
using the formalism presented in \cite{benanew}.  One can observe
quite a few features similar to the ones depicted in
Fig.~\ref{Fig1}C, including the existence of a three-fold
symmetry. However the three-fold anisotropy is much less
pronounced.

\begin{figure}
\begin{center}
\epsfig{file=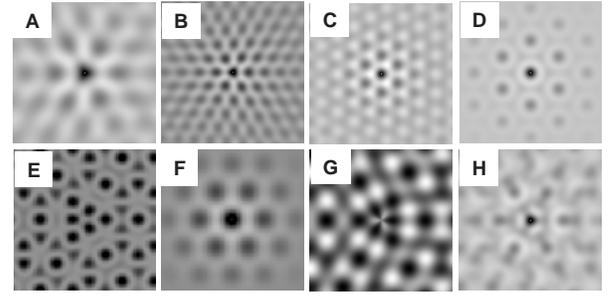,width=8cm,angle=0}
\end{center}
\caption{Calculated real-space LDOS modulations in a monolayer
graphene for a point defect placed on top of an A atom.
Figs.~\ref{Fig2} A, B, C depict the contributions of selected
scattering processes to the LDOS ($K \rightarrow K_{1}, K
\rightarrow K_{2}$ and $K \rightarrow K_{3}$ respectively) as
indicated in Fig.~\ref{Fig1}D; in \ref{Fig2}F, and \ref{Fig2}G we
depict the separate contributions of the A and B sublattices to
the $K\rightarrow K_1$ LDOS modulations in \ref{Fig2}A.
Figs.~\ref{Fig2}D and E depict the A and B sublattice
contributions to the full LDOS, evaluated when all scattering
processes are considered. In H we depict the sum of the A and B
contributions to the full LDOS, when the weight of the
$K\rightarrow K_1$  B component is increased three times compared
to that of the $K\rightarrow K_1$ A component.}\label{Fig2}
\end{figure}

Figure \ref{Fig2} displays the calculated real-space modulations
in the LDOS close to the point defect, when only specific
scattering processes have been considered. For simplicity this
calculation has been performed for a monolayer graphene, as for
bilayer graphene it is less clear how much the two bands and the
two layers contribute to the measured LDOS. The intra-nodal
scattering processes are responsible for radially-symmetric
features with small $\overrightarrow{q}$ wavevectors and should
therefore not be responsible for the presence of the three-fold
symmetry. Hence we do not focus on these modulations here.

In Figs.~\ref{Fig2}A, \ref{Fig2}B and \ref{Fig2}C, we plot the
LDOS modulations due solely to scattering between nearest,
second-nearest, and third-nearest neighboring nodes, generically
denoted $K \rightarrow K_1$, $K \rightarrow K_2$ and $K
\rightarrow K_3$ respectively, as schematized in Fig.~\ref{Fig1}D.
First, it is clear that the most important effect of the impurity,
which is responsible for the $R3$ patterns observed
experimentally, is the $K \rightarrow K_1$ scattering. Second,
close to the impurity, the patterns are almost six-fold symmetric,
with a slight three-fold anisotropy. Thus, none of these
scattering events contributes significantly to the threefold
symmetry. To investigate the source of the anisotropy, we first
note that, in Ref.\cite{BenaPRL08} it was shown that the  $K
\rightarrow K_1$ high-intensity regions appearing in the FT of the
LDOS at the corners of the BZ are anisotropic, more pronouncedly
for monolayer graphene than for bilayer graphene. This is due to
the chirality of the quasiparticles in graphene. We believe that
this asymmetry in the observed momentum space features, and thus
the chirality of quasiparticles, is responsible for the small
amount of $C3v$ asymmetry observed in the theoretical real-space
LDOS calculations. The second observation is that the anisotropy
comes mainly from the B component of the LDOS, that describes the
double virtual hopping of quasiparticles between the A and B
sublattices\cite{ab}. This is detailed in Figs.~\ref{Fig2}F and
\ref{Fig2}G, where we plot separately the A and the B sublattice
contributions to the LDOS modulations generated by $K \rightarrow
K_1$ scattering. We can see that the contribution from the A atoms
is fully six-fold symmetric, while the contribution of the B atoms
is three-fold symmetric. This is also true for the A and B
components (depicted in \ref{Fig2}D and \ref{Fig2}E respectively)
of the full LDOS obtained from summing the contributions of all
scattering processes. Fig.~\ref{Fig2}H depicts the sum of the A
and B contributions to the full LDOS, where in order to increase
the anisotropy, we have multiplied the $K\rightarrow K_1$ B
contribution by three. This may mimic bilayer graphene where, due
to the coupling between the two layers, the sublattices A and B
may not contribute equally to the observed LDOS. A very strong
threefold symmetry is observed in this weighted superposition, and
the resulting image corresponds more closely to the pattern
observed experimentally in figure \ref{Fig1}B). This is consistent
with having a defect at an A site whose dominant effect is the
scattering of sublattice B electrons between two consecutive
valleys ($K \rightarrow K_1$).

\begin{figure}
\begin{center}
\epsfig{file=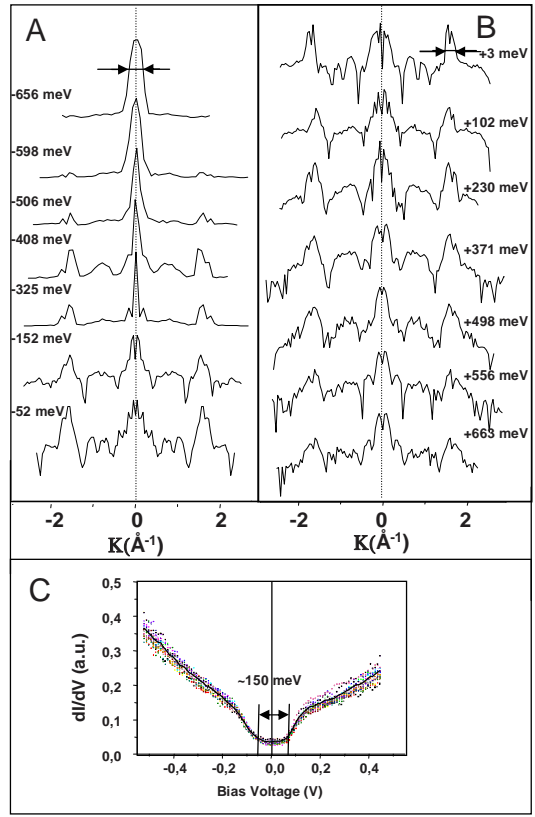,width=7cm,angle=0}
\end{center}
\caption{A) and B) Intensity profiles along the $K-\Gamma-K$
direction of the 2D FT of the LDOS taken for different bias
voltages. C) STS measurement taken near the impurity showing a gap
centered around the Fermi level. }\label{Fig3}
\end{figure}

\begin{figure}
\begin{center}
\epsfig{file=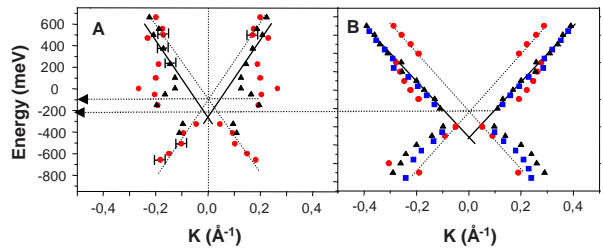,width=8cm,angle=0}
\end{center}
\caption{The dispersion for the width of the central-ring feature
(red circles) and of the K-points feature (black triangles)
obtained from the experimental data (A), and from the theoretical
$K-\Gamma-K$ cuts (B). The dispersion of the central-feature
shoulder on the theoretical curves is indicated by blue
squares.}\label{Fig4}
\end{figure}

We are also interested in how such a defect affects the
relativistic character of the quasiparticles. To find this, we
measured the quasiparticle energy dispersion using the Fourier
transform of conductance images at different energies. Every image
was acquired using a lock-in amplifier and a modulation voltage of
$\pm 20$ $meV$ which gave the energy uncertainty. We have
performed a 2D FFT of these images, and in Figs.~\ref{Fig3}A and
\ref{Fig3}B we show the resulting intensity profiles along the
direction $K-\Gamma-K$. For a wide range of energies, between
energies well below the Fermi level ($-950$ $meV$) to about $-150
meV$, the width of the peak at the center of the Brillouin zone
(measured at half maximum) shows a clear dispersion with the bias
voltage (it decreases when the energy moves towards the Fermi
level). At energies closer to the Fermi level the profile becomes
more complex and displays a dip in the intensity profile close to
the $\Gamma$ point. The central peak also shows two lateral
structures (shoulders). The width of the central feature does not
appear to disperse with energy close to the Dirac point, but it
starts dispersing again for higher positive energies. In
Fig.~\ref{Fig3}C we show the STS measurement of the LDOS ($dI/dV$
as a function of bias voltage) taken close to the impurity. We can
see a $150$ $meV$ gap-like feature centered around the Fermi level
at $0$ $meV$.
Fig.~\ref{Fig4} shows a linear dependence of the width of the
central feature with energy, except in an energy range from $-200$
$meV$ to $100$ $meV$. For these energies the shape of the central
peak is more complex. The width of the features at the K points
follows a similar dispersion. Beyond $+500$ $meV$, all the
features seem to follow a different dispersion branch, as
indicated in Fig.~\ref{Fig4}. We estimate the Dirac point at
$-100$ $meV$ below the Fermi level. The spreading in the
dispersion of the points close to the Fermi level could be
attributed to the presence of the gap. We compare the experimental
results with similar theoretical profiles obtained using a
T-matrix approximation for a single localized impurity in bilayer
graphene \cite{BenaPRL08}. No gap at the Fermi level was included
in the theoretical calculation, while a small gap of $\approx 100$
$meV$ was assumed near the Dirac point, which was taken to be
close to $-250$ $meV$. In Fig.~\ref{Fig4}B we plot the dispersion
of the central and K-points features, and of the shoulder,
obtained from the theoretical curves. We also observe the presence
of two different dispersion branches. The second dispersion branch
and the shoulder arise because of the bilayer graphene upper band
which opens at energies higher than the inter-layer coupling. For
monolayer graphene these extra features should not appear.

In conclusion, our measurements show that the quasiparticle
approximation and the Fermi liquid theory are robust over a large
range of energies. The complex structure of the central feature
(the existence of the shoulder), as well as the presence of two
dispersion branches are consistent with the bilayer (or
multilayer) character of the graphene sample. While the point
defect modifies the electronic wave-function in its vicinity, a
clear linear dispersion is still observed, and the relativistic
character of the quasiparticle is preserved. Also, the STS
measurements in Fig.~\ref{Fig3}C indicate the presence of a gap
centered at the Fermi point, and not at the Dirac point inferred
from our FT-STM measurements. This gap is thus different from the
Dirac-point gap observed by ARPES
\cite{BostwickNaturephysics07,ZhouNaturePhysics07} which was
attributed to the different doping levels of the epitaxial
graphene layers. It is however consistent with previous STM
measurements performed on epitaxial graphene\cite{BrarAPL07}, and
more recently on exfoliated graphene \cite{Crommie08}, where a gap
observed at the Fermi level was attributed to a pinning of the
tunneling spectrum due to the coupling with phonons.

A marked three-fold symmetry in the vicinity of the point defect
was also observed. Detailed theoretical studies of the different
scattering events underline the importance of the scattering
between consecutive nodal points ($K \rightarrow K_1$). Our
analysis also suggests that for bilayer graphene, the two
sublattices do not contribute equally to the measured LDOS. Thus,
the amount of asymmetry observed in the LDOS near the impurity is
consistent with the fact that the scattering of the electrons
between the sites of the sublattice on which the impurity is
located (A) and the electrons on the B sublattice dominate over
the scattering of the electrons between sites of the same (A)
sublattice.

\acknowledgments We would like to thank P. Y. Clement for useful
discussions. This work is supported by the R\'{e}gion Alsace and
the CNRS. C.~Bena acknowledges the support of a Marie Curie Action
under FP6.

\newpage
\end{document}